\documentstyle[12pt]{article}
\title{
A program for coupled-channels calculations with all order couplings for
heavy-ion fusion reactions \\}
\author{K. Hagino$^{1}$, N. Rowley$^{2}$,
and A.T. Kruppa$^{3}$
\\ \\
\medskip
{\it $^{1}$ Institute for Nuclear Theory, Department of Physics, } \\
{\it University of Washington, Seattle, WA 98195, USA}
\\
{\it $^{2}$ Institute de Recherches Subatomiques (IReS), 23 rue du Loess, }\\
{\it F--67037 Strasbourg Cedex 2, France}
\\
{\it $^{3}$ Institute of Nuclear Research of the Hungarian Academy of
Science,} \\
{\it Pf. 51, H--4001 Debrecen, Hungary}
}
\date{}

\begin{document}

\maketitle

\begin{center}
{\bf Abstract}
\end{center}

A FORTRAN77 program is presented that calculates fusion cross sections
and mean angular momenta of the compound nucleus under the influence of
couplings between the relative motion and several nuclear collective
motions.
The no-Coriolis approximation is employed to reduce
the dimension of coupled-channels equations.
The program takes into account the effects of non-linear
couplings to all orders, which have been shown to play an important
role in heavy-ion fusion reactions at subbarrier energies.

\bigskip


\newpage

\begin{center}
{\bf PROGRAM SUMMARY}
\end{center}

\noindent
{\it Title of program:} CCFULL \\

\noindent
{\it Catalogue identifier:} 17.7 (Experimental
Analysis - Fission, Fusion, Heavy-ion)\\

\noindent
{\it Distribution format:} ASCII \\

\noindent
{\it Computer for which the program is designed and others on which
it has been tested:} any UNIX work-station or PC. The program has been
tested on DEC and DEC-Alpha. \\



\noindent
{\it Operating system or monitor under which the program has been tested:}
UNIX \\

\noindent
{\it Programming language used:} FORTRAN 77\\







\noindent
{\it Nature of physical problem}

\noindent
It has by now been well established that fusion reactions at energies
near and below the Coulomb barrier are strongly influenced by couplings
of the relative motion of the colliding nuclei to several nuclear intrinsic
motions. Recently, precisely measured fusion cross sections have become
available for several systems, and a distribution of the Coulomb barrier,
which is originated from the channel couplings, have been extracted.
It has been pointed out that the linear coupling approximation,
which has often been used in coupled-channels calculations, is inadequate
in order to analyze such high presicion experimental data.
The program CCFULL solves the coupled-channels equations to compute
fusion cross sections
and mean angular momenta of compound nucleus, taking into account the
couplings to all orders. \\

\noindent
{\it Method of solution}

\noindent
CCFULL directly integrates coupled second order differential equations
using the modified Numerov method.
The incoming wave boundary condition is employed and a barrier penetrability
is calculated for each partial wave.
Nuclear coupling matrix elements are evaluated by using the matrix
diagonalisation method once the physical space has been defined. \\

\noindent
{\it Restrictions on the complexity of the program}

\noindent
The program is best suited for systems where the number of
channels which strongly couple to the ground state is relatively small
and where multi-nucleon transfer reactions play less important role compared
with inelastic channels. It also relies on an assumption that the fusion
process is predominantly governed by quantum
tunneling over the Coulomb barrier.
This assumption restricts a system which the program can handle to
that where the sum of the charge of the projectile and the target nuclei
$Z_p+Z_T$ is larger than around 12 and the charge product $Z_pZ_T$
less than around 1800.
For most of experimental data which were measured to aim to extract
fusion barrier distributions, this condition is well satisfied.
The program also treats a vibrational coupling in the harmonic
limit and a rotational coupling with a pure rotor. The program can be
modified for general couplings by directly providing coupling strengths and
excitation energies. \\

\noindent
{\it Typical running time}\\
\noindent
A few seconds for input provided. The computer time
depends strongly upon the number of channels to be included.
It will considerably
increase if one wishes to include a large number of channels, as for
instance 20.



\newpage

\begin{center}
{\bf LONG WRITE-UP}
\end{center}

\noindent
{\bf 1. Introduction}

Fusion is defined as a reaction where two separate nuclei combine together
to form a composite system.
When the incident energy is not so large and the system is not so
light, the reaction process
is predominantly governed by quantum
tunneling over the Coulomb barrier created by the strong cancellation
between the repulsive Coulomb force and the attractive nuclear
interaction.
Extensive experimental as well as theoretical studies have revealed
that fusion reactions at energies near and below the Coulomb barrier
are strongly influenced by
couplings of the relative motion of the
colliding nuclei to several nuclear intrinsic motions \cite{B88}.
Heavy-ion subbarrier fusion reactions thus provide a good
opportunity to address the general problem on quantum tunneling in the
presence of couplings, which has been a popular subject in
the past decade in many branches of physics and chemistry.

Thanks to the recent developments in experimental techniques, fusion
cross sections can now become measured with high accuracy in small
energy intervals.
Such high precision experimental data
have generated a renewed interest in heavy-ion subbarrier
fusion reactions in recent years \cite{DHRS98,BT98}.
For instance, they have enabled
a detailed study of the effects of couplings on fusion reactions through
the so called fusion barrier distribution \cite{RSS91,LDH95}
and have thus offered a good opportunity to test any theoretical
framework for subbarrier fusion reactions.

Theoretically the standard way to address the effects of the
coupling between
the relative motion and the intrinsic degrees of freedom on fusion is to
numerically solve the coupled-channels equations, including all the
relevant channels.
In the past, the coupled-channels calculations were often performed
using the linear coupling approximation, where the
coupling potential is expanded in powers of the deformation
parameter, keeping only the linear term.
It has been demonstrated that non-linear couplings significantly affect
the shape of fusion barrier distributions and thus the linear coupling
approximation is inadequate in quantitative comparison with the recent
high quality data of fusion cross sections \cite{HTD97,SACN95}.
The program CCFULL includes the couplings to full order and thus it does not
introduce the expansion of the coupling potential.
Since the dimension of the coupled-channels equations with full space
is in general too large for practical purposes, the program employs the
no-Coriolis approximation, which is sometimes referred to as
the isocentrifugal
approximation too, to reduce the dimension \cite{HTBB95,LR84}.
For heavy-ion fusion reactions, this approximation has been confirmed to
work well \cite{T87}.
The program is otherwise exact and take full account of the finite excitation
energies of intrinsic motions. It includes Coulomb excitations and
uses the incoming wave boundary condition inside the Coulomb barrier.

\medskip

\noindent
{\bf 2. Coupled-channels equations}

For heavy-ion fusion reactions, to a good approximation
one can replace the angular momentum of the relative motion in
each channel by the total angular momentum $J$\cite{HTBB95,LR84}.
This approximation, often referred to as
no-Coriolis approximation or isocentrifugal approximation, is used
in the program. The coupled-channels equations then read
\begin{equation}
\left[-\frac{\hbar^2}{2\mu}\frac{d^2}{dr^2}
+\frac{J(J+1)\hbar^2}{2\mu r^2}
+V_N^{(0)}(r)+\frac{Z_PZ_Te^2}{r}+\epsilon_n
-E\right]\psi_n(r)+\sum_mV_{nm}(r)\psi_m(r)=0  ,
\label{cc}
\end{equation}
where $r$ is the radial component of the coordinate of the
relative motion and $\mu$ is the reduced mass, respectively.
$E$ is the
bombarding energy in the center of mass frame and
$\epsilon_n$ is the excitation energy of the $n$-th channel.
$V_{nm}$ are the matrix elements of the coupling Hamiltonian,
which in the collective model
consist of Coulomb and nuclear components. These two components are
detailed in the following section.
$V_N^{(0)}$ is the nuclear potential in the entrance channel.
In the program, the Woods-Saxon parametrisation
\begin{equation}
V_N^{(0)}(r)= -\frac{V_0}{1+\exp((r-R_0)/a)};
~~~R_0=r_0\left(A_P^{1/3}+A_T^{1/3}\right),
\label{vn}
\end{equation}
is adopted for the nuclear potential $V_N^{(0)}$.

The coupled-channels equations are solved
by imposing the boundary conditions that
there are only incoming waves at $r=r_{min}$, and there are only
outgoing waves at infinity for all channels except the entrance
channel ($n$=0), which has an incoming wave with amplitude one as well.
This boundary condition is referred to as the incoming wave
boundary condition (IWBC) \cite{LP84}, and is valid for
heavy-ion reactions, where there is a strong absorption inside the
Coulomb barrier.
The program CCFULL adopts the minimum position of the Coulomb pocket inside
the barrier for $r_{min}$.
Practically the numerical solution is matched
to a linear combination of incoming and outgoing and Coulomb wave
functions at finite distance $r_{max}$ beyond which both the nuclear
potential and the Coulomb coupling are sufficiently small.
The boundary conditions are thus expressed as
\begin{eqnarray}
\psi_n(r)&\to& T_n\exp\left(-i\int^r_{r_{min}}k_n(r')dr'\right)
~~~~~~~~~~~~~~~r \le r_{min}, \label{bc1}\\
&\to&
H_J^{(-)}(k_nr) \delta_{n,0} + R_nH_J^{(+)}(k_nr)~~~~~~~~~~~~~r>r_{max},
\label{bc2}
\end{eqnarray}
where
\begin{equation}
k_n(r)=\sqrt{\frac{2\mu}{\hbar^2}\left(E-\epsilon_n
-\frac{J(J+1)\hbar^2}{2\mu r^2}
-V_N(r)-\frac{Z_PZ_Te^2}{r}-V_{nn}(r)\right)},
\end{equation}
is the local wave number for the $n$-th channel and
$k_n=k_n(r=\infty)$.
$H_J^{(-)}$ and $H_J^{(+)}$ in eq. (\ref{bc2}) is the incoming and the
outgoing Coulomb functions, respectively.

In order to ensure that there are only incoming waves at
$r\to r_{min}$, the program CCFULL solves the coupled-channels
equations outwards from $r_{min}$, first by setting \cite{RMR78}
\begin{eqnarray}
\psi_n(r_{min})&=&1,
~~~~~~~~~~~~~~~~~~~~\psi_m(r_{min})=0~~(m\neq n), \\
\frac{d}{dr}\psi_n(r_{min})&=&-ik_n(r_{min}),
~~~~\frac{d}{dr}\psi_m(r_{min})=0~~(m\neq n).
\end{eqnarray}
Since the first derivative of the wave functions at $r_{min}$ has been
explicitly written down from eq. (\ref{bc1}),
the wave functions at $r=r_{min} + h$, $h$ being the radial
mesh to integrate the equations, can be determined
in the Runge-Kutta method.
After the wave functions at $r=r_{min}+h$ have been thus obtained,
CCFULL solves the coupled-channels equations from $r=r_{min} + h$
to $r=r_{max}$ in the modified Numerov methods \cite{MSR66}, since
the Runge-Kutta method may not be so efficient to solve
the second order differential equations.
The modified Numerov method relates the wave functions at
$r_{i+1}\equiv r_{min}+(i+1)h$ to those at
$r_i$ and $r_{i-1}$ as
\begin{equation}
\vec{\psi}^{i+1}=\left(1-\frac{h^2}{12}{\cal A}^{i+1}\right)^{-1}
\left[\left\{\left(\frac{h^2}{\sqrt{12}}{\cal A}^{i}+\sqrt{3}\right)^2
-1\right\}
\left(1-\frac{h^2}{12}{\cal A}^i\right)\vec{\psi}^i
-\left(1-\frac{h^2}{12}{\cal A}^{i-1}\right)\vec{\psi}^{i-1}\right],
\end{equation}
where ${\cal A}_{nm}(r)$ is defined by
\begin{equation}
{\cal A}_{nm}(r)=\frac{2\mu}{\hbar^2}
\left[\left(V_N^{(0)}(r)+\frac{J(J+1)\hbar^2}{2\mu r^2}
+\frac{Z_PZ_Te^2}{r}+\epsilon_n-E\right)\delta_{n,m}
-V_{nm}(r)\right],
\end{equation}
and $\vec{\psi}^{i}$ are the wave functions at $r_i$.

Let $\chi_{nm}(r)$ be the wave function of the $m$-th channel
thus obtained, i.e. it is $\psi_m(r)$ which satisfies the boundary
conditions eq. (\ref{bc1}) at $r=r_{min}$.
At $r=r_{max}$, $\chi_{nm}$ can be expressed by a
superposition of the incoming and outgoing Coulomb waves as
\begin{equation}
\chi_{nm}(r)=C_{nm}H^{(-)}_J(k_mr) + D_{nm}H^{(+)}_J(k_mr)
~~~~~~~~r\to r_{max}.
\end{equation}
The coefficients $C_{nm}$ and $D_{nm}$ are determined
either by matching the logarithmic derivatives at $r_{max}$ or by
matching the ratio of the wave functions at $r_{max}-h$
to those at $r_{max}+h$. Since the modified
Numerov methods does not automatically generate the derivative of
the wave functions, the latter procedure is more suitable here.
The coefficients are then obtained as
\begin{equation}
C_{nm}=\frac{H^{(+)(i-1)}_{Jm}\chi_{nm}^{(i+1)}
-H^{(+)(i+1)}_{Jm}\chi_{nm}^{(i-1)}}
{H^{(+)(i-1)}_{Jm}H^{(-)(i+1)}_{Jm}-H^{(+)(i+1)}_{Jm}H^{(-)(i-1)}_{Jm}}
\end{equation}
and
\begin{equation}
D_{nm}=\frac{H^{(-)(i-1)}_{Jm}\chi_{nm}^{(i+1)}
-H^{(-)(i+1)}_{Jm}\chi_{nm}^{(i-1)}}
{H^{(-)(i-1)}_{Jm}H^{(+)(i+1)}_{Jm}-H^{(-)(i+1)}_{Jm}H^{(+)(i-1)}_{Jm}}
\end{equation}
respectively. We have defined
$H^{(+)(i+1)}_{Jm}\equiv H^{(+)}_J(k_m\cdot(r_{max}+h))$, etc. and
$\chi_{nm}^{i+1}\equiv\chi_{nm}(r_{max}+h)$, etc.
This procedure is repeated for all $n$ and $m$
to determine the matrices $C$ and $D$.

The solution of the coupled-channels equations
with the proper boundary conditions (\ref{bc1}) and (\ref{bc2})
is given by a linear combination of $\chi_{nm}$ as
\begin{equation}
\psi_m(r)=\sum_n T_n\chi_{nm}(r).
\end{equation}
This equation satisfies the boundary condition (\ref{bc1}) at
$r=r_{min}$. At $r=r_{max}$, it leads to
\begin{equation}
\psi_m(r_{max})=\sum_n T_n\chi_{nm}(r_{max})
=\sum_n T_n
\left(C_{nm}H^{(-)}_J(k_mr_{max}) + D_{nm}H^{(+)}_J(k_mr_{max}) \right).
\label{match}
\end{equation}
By comparing between eqs. (\ref{bc2}) and (\ref{match}), one finds
\begin{equation}
\sum_n T_nC_{nm}=\delta_{m,0}.
\end{equation}
The transmission coefficients are then finally obtained by
\begin{equation}
T_n=\left(C^{-1}\right)_{n0}.
\end{equation}
For many examples, we are interested only in the inclusive
process, where the intrinsic degree of freedom emerges
in any final state.
Taking a summation over all possible intrinsic states,
the inclusive penetrability is given by
\begin{equation}
P_J(E)=\sum_n\frac{k_n(r_{min})}{k_0}\left|T_n\right|^2.
\end{equation}
The fusion cross section and the mean angular momentum of compound nucleus
are then calculated by
\begin{eqnarray}
\sigma_{fus}(E)&=&\sum_J\sigma_J(E)=\frac{\pi}{k_0^2}\sum_J(2J+1)P_J(E),\\
<l>&=&\sum_JJ\sigma_J(E)/\sum_J\sigma_J(E), \nonumber \\
&=&\left(\frac{\pi}{k_0^2}\sum_JJ(2J+1)P_J(E)\right)\left/
\left(\frac{\pi}{k_0^2}\sum_J(2J+1)P_J(E)\right)\right.,
\end{eqnarray}
respectively. In the program CCFULL, the summation over the partial
wave is truncated at the angular momentum whose contribution
to the cross section is less than 10$^{-4}$ times total cross section.

\medskip

\noindent
{\bf 3. Coupling matrix elements}

\medskip

\noindent
{\bf 3.1. Rotational coupling}

In this section, we give explicit expressions for the coupling matrix
elements $V_{nm}(r)$ in eq. (\ref{cc}).
Let us first consider a rotational coupling in the target nucleus.
The nuclear coupling Hamiltonian can be generated by changing the
target radius in the nuclear potential (\ref{vn}) to a dynamical
operator
\begin{equation}
R_0 \to R_0 + \hat{O}= R_0+\beta_2R_T Y_{20}
+ \beta_4 R_TY_{40},
\end{equation}
where $R_T$ is parametrised as $r_{coup}A_T^{1/3}$, and
$\beta_2$ and $\beta_4$ are the quadrapole and hexadecapole
deformation parameters of the deformed target nucleus, respectively.
The nuclear coupling Hamiltonian is thus given by
\begin{equation}
V_N(r,\hat{O})= -\frac{V_0}{1+\exp((r-R_0-\hat{O})/a)}.
\label{Ncoup}
\end{equation}
We need matrix elements of this coupling Hamiltonian between
the $|n>=|I0>$ and $|m>=|I'0>$ states of the ground rotational band of the
target. These can be easily obtained using a matrix algebra \cite{KR93}.
In this algebra, one first looks for the eigenvalues and eigenvectors
of the operator $\hat{O}$ which satisfies
\begin{equation}
\hat{O}|\alpha> = \lambda_{\alpha} |\alpha>.
\end{equation}
In the program CCFULL, this is done by diagonalising the matrix
$\hat{O}$, whose elements are given by
\begin{eqnarray}
\hat{O}_{II'}
&=&
\sqrt{\frac{5(2I+1)(2I'+1)}{4\pi}}\beta_2 R_T
\left(\begin{array}{ccc}
I&2&I'\\
0&0&0
\end{array}\right)^2 \nonumber \\
&&+
\sqrt{\frac{9(2I+1)(2I'+1)}{4\pi}}\beta_4 R_T
\left(\begin{array}{ccc}
I&4&I'\\
0&0&0
\end{array}\right)^2.
\end{eqnarray}
The nuclear coupling matrix elements are then evaluated as
\begin{eqnarray}
V_{nm}^{(N)}&=&<I0|V_N(r,\hat{O})|I'0>-V_N^{(0)}(r)\delta_{n,m}, \nonumber \\
&=&\sum_{\alpha}<I0|\alpha><\alpha|I'0>V_N(r,\lambda_{\alpha})
-V_N^{(0)}(r)\delta_{n,m}.
\label{full}
\end{eqnarray}
The last term in this equation is included to avoid the double counting
of the diagonal component.

For the Coulomb interaction of the deformed target, the program CCFULL
includes up to the second order with respect to $\beta_2$ and to the
first order of $\beta_4$. Contrary to the nuclear couplings, the higher
order couplings of the Coulomb interaction have been shown to play a rather
minor role \cite{HTD97}. The matrix elements are then given by
\begin{eqnarray}
V_{nm}^{(C)}&=&
\frac{3Z_PZ_T}{5}\frac{R_T^2}{r^3}
\sqrt{\frac{5(2I+1)(2I'+1)}{4\pi}}
\left(\beta_2+\frac{2}{7}\sqrt{\frac{5}{\pi}}\beta_2^2\right)
\left(\begin{array}{ccc}
I&2&I'\\
0&0&0
\end{array}\right)^2 \nonumber \\
&&+
\frac{3Z_PZ_T}{9}\frac{R_T^4}{r^5}
\sqrt{\frac{9(2I+1)(2I'+1)}{4\pi}}
\left(\beta_4+\frac{9}{7}\beta_2^2\right)
\left(\begin{array}{ccc}
I&4&I'\\
0&0&0
\end{array}\right)^2.
\end{eqnarray}
The total coupling matrix element is given by the sum of $V_{nm}^{(N)}$
and $V_{nm}^{(C)}$.

\medskip

\noindent
{\bf 3.2. Vibrational coupling}

We next consider a vibrational coupling. Ref. \cite{HTD97} discusses
all order nuclear couplings for the case where the vibration can be
approximated by the harmonic oscillator.
In realistic case, however, phonon spectra are often truncated at some
level, and thus the intrinsic motion deviates from the harmonic limit
even when the excitation energies are equal spaced and/or the electro magnetic
transitions do not alter in the linear approximation. (See ref. \cite{BBH99}
for a discussion on differences between the harmonic oscillator and the
truncated oscillator, i.e. spin systems.) In such a situation, the matrix
formalism discussed in the previous section still
provides a convenient and powerful technique to evaluate the coupling
matrix elements \cite{DHRS98}.
For vibrational coupling, the operator $\hat{O}$ in the nuclear coupling
Hamiltonian is given by
\begin{equation}
\hat{O}=\frac{\beta_{\lambda}}{\sqrt{4\pi}}R_T
(a_{\lambda 0}^{\dagger} + a_{\lambda 0}),
\end{equation}
where $\lambda$ is the multipolarity of the vibrational mode and
$a_{\lambda 0}^{\dagger} (a_{\lambda 0})$ is the creation (annihilation)
operator of the phonon.
The matrix element of this operator between the $n$-phonon state $|n>$ and
the $m$-phonon state $|m>$ is given by
\begin{equation}
\hat{O}_{nm}=\frac{\beta_{\lambda}}{\sqrt{4\pi}}R_T
(\sqrt{m}\delta_{n,m-1}+\sqrt{n}\delta_{n,m+1}).
\end{equation}
The rest of the procedure to evaluate the nuclear coupling matrix element
is exactly the same as the rotational case.
The operator $\hat{O}$ is diagonalised in a physical space and then
the nuclear coupling matrix elements are calculated according to
eq. (\ref{full}).

The program CCFULL uses the linear coupling approximation for the Coulomb
coupling of the vibrational degree of freedom.
The Coulomb coupling matrix elements are thus read
\begin{equation}
V^{(C)}_{nm}(r)=\frac{\beta_{\lambda}}{\sqrt{4\pi}}
\frac{3}{2\lambda+1}Z_PZ_Te^2
\frac{R_T^{\lambda}}{r^{\lambda+1}}
(\sqrt{m}\delta_{n,m-1}+\sqrt{n}\delta_{n,m+1}).
\end{equation}
Again the total coupling matrix element is given by the sum of $V_{nm}^{(N)}$
and $V_{nm}^{(C)}$.

\medskip

\noindent
{\bf 3.3. Transfer coupling}

The program CCFULL includes a pair-transfer coupling between the ground
states. It uses the macroscopic coupling form factor given by \cite{DV86}
\begin{equation}
F_{trans}(r)=F_t \frac{dV_N^{(0)}}{dr},
\label{trans}
\end{equation}
where $F_t$ is the coupling strength.

\medskip

\noindent
{\bf 4. Program input and test run}

A description of the format for the input parameters is given
in table 1. All parameters are entered in free format.
The first line contains the parameters specifying the system.
AP (AT) is the projectile (target) mass and ZP (ZT) is the projectile
(target) charge.
The second line is for the coupling Hamiltonian. RP (RT) is the
radius parameter $r_{coup}$
of the projectile (target) used in the coupling Hamiltonian.
Note that this is in general different from the radius parameter used in the
nuclear potential (\ref{vn}), which is defined in the seventh line.
IVIBROTP (IVIBROTT) is an option which specifies the property
of the intrinsic motion of the projectile (target). If it is set to be
$-1$, the projectile (target) is assume to be inert and the
fifth (the third and the fourth) line will be ignored.
The fusion cross sections and the mean angular momentum in the absence
of the channel coupling can be therefore obtained by setting both
the IVIBROTP and the IVIBROTT to be $-1$.
When IVIBROTP (IVIBROTT) is set to be zero,
the CCFULL assumes that the coupling in the projectile (target)
is vibrational, while if it is
set to be one, the rotational coupling is assumed.

The third line is for detailed information on the target excitation.
If IVIBROT is zero (i.e., the vibrational coupling), the CCFULL reads
OMEGAT, BETAT, LAMBDAT, and NPHONONT.
OMEGAT is the excitation energy of the single phonon state, BETAT is the
deformation parameter, and LAMBDAT is the multipolarity of the vibrational
excitation. NPHONONT is the maximum phonon number to be included. For example,
if it is two, up to two phonon states are included in the calculation.
The CCFULL assumes the harmonic oscillator for a vibrational coupling. The
excitation energy of the $n$-phonon state is thus given by $n$ times
OMEGAT. Sometimes a user may want to use a different value of deformation
parameter for the nuclear coupling from that for the Coulomb coupling.
The CCFULL therefore will ask a user interactively before a run
whether he/she intends to use a
different value of the coupling strength for the nuclear coupling.
If IVIBROTT is one (i.e.,
the rotational coupling), the CCFULL reads E2T, BETA2T, BETA4T, and NROTT.
E2T is the excitation energy of the first 2$^+$ state in the ground rotational
band of the target nucleus, BETA2T and BETA4T are the quadrapole and
hexadecapole deformation parameters, respectively. NROTT is the number of
levels in the rotational band to be included. For instance, if it
is 3, the 2$^+$, 4$^+$
and 6$^+$ states are included together with the ground state. The CCFULL
assumes a pure rotor for a deformed nucleus, and the excitation energy of the
$I^+$ state is given by $I(I+1)\cdot$E2T/6.

In many applications, there are two vibrational modes of excitations in the
target nucleus. A typical example is the octupole and quadrapole
vibrational excitations
in $^{144}$Sm. The fourth line is for the second mode of excitation in the
target nucleus. The meaning of OMEGAT2, BETAT2, LAMBDAT2 and NPHONONT2 is
the same as OMEGAT, BETAT, LAMBDAT and NPHONONT, respectively.
The second mode is not included when NPHONONT2 is set to be zero.
OMEGAT2, BETAT2, and LAMBDAT2 are then ignored.
When NPHONONT2 is not zero, a user will be asked before a run which of the
mutual excitation channels he/she intends to include in the calculation.

The fifth line is the same as the third line, but for the projectile
excitations. If there exist excitations both in the projectile and the
target, the CCFULL takes into account all the possible mutual excitation
channels between the projectile and the target excitations.

The sixth line is for the pair transfer coupling. QTRANS is the Q-value for
the pair transfer channel, while FTR is the coupling strength defined by
eq. (\ref{trans}).
NTRANS is the number of the pair transfer channel. In the present version
of the CCFULL, NTRANS is restricted to be either one or zero. If it is zero,
the pair transfer channel is not included and QTRANS and FTR are ignored.

The seventh line is for the nuclear potential in the entrance
channel (\ref{vn}). V0 is the depth parameter of the Woods-Saxon
potential, R0 is the radius parameter $r_0$ in eq. (\ref{vn}), and A0
is the surface diffuseness parameter $a$.

EMIN, EMAX, and DE in the next line are the minimum and the maximum
value of the colliding energy in the center of mass frame and the
interval in the energy scale, respectively. The CCFULL constructs the
distribution of partial cross sections $\sigma_J$ as a function of $J$
if a single value of the energy is entered, i.e. either when
EMIN=EMAX or DE=0.

The accuracy of the calculation is controlled by the matching radius
RMAX and the mesh for the integration DR in the ninth line.
For many application, especially for asymmetric system such as
$^{16}$O + $^{144}$Sm, RMAX=30 fm and DR=0.05 fm provides sufficiently
accurate results. For heavier systems, such as $^{64}$Ni + $^{92}$Zr,
RMAX may have to be extended as large as 50 fm.

The test case shows the fusion cross sections and the mean angular momentum
of the compound nucleus for the $^{16}$O + $^{144}$Sm reaction. The projectile
nucleus $^{16}$O is assumed to be inert, while the single octupole
phonon excitation in $^{144}$Sm is taken into account. The transfer channel
is not included in this calculation.

\bigskip

\noindent
{\bf Acknowledgment}

The authors thank M. Dasgupta for discussions.
K.H. acknowledges support from the Japan Society for the Promotion
of Science for Young Scientists.

\newpage

\begin{table}
\caption{Input to the computer code CCFULL.}
\begin{center}
\begin{tabular}{|l|l|}
\hline
Line 1 & AP, ZP, AT, ZT \\
\hline
Line 2 & RP, IVIBROTP, RT, IVIBROTT \\
\hline
Line 3 & OMEGAT, BETAT, LAMBDAT, NPHONONT  (if IVIBROTT=0) \\
       & E2T, BETA2T, BETA4T, NROTT~~~~~~~~~~~~~~~~~~(if IVIBROTT=1) \\
\hline
Line 4 & OMEGAT2, BETAT2, LAMBDAT2, NPHONONT2 \\
\hline
Line 5 & OMEGAP, BETAP, LAMBDAP, NPHONONP  (if IVIBROTP=0) \\
       & E2P, BETA2P, BETA4P, NROTP~~~~~~~~~~~~~~~~~~(if IVIBROTP=1) \\
\hline
Line 6 & NTRANS, QTRANS, FTR \\
\hline
Line 7 & V0, R0, A0 \\
\hline
Line 8 & EMIN, EMAX, DE \\
\hline
Line 9 & RMAX, DR \\
 \hline
\end{tabular}
\end{center}
\end{table}

\newpage

\bigskip

\begin{center}
{\bf TEST RUN INPUT}
\end{center}

\begin{verbatim}

16.,8.,144.,62.
1.2,-1,1.06,0
1.81,0.205,3,1
1.66,0.11,2,0
6.13,0.733,3,0
0,0.,0.3
105.1,1.1,0.75
55.,72.,1.
30,0.05

\end{verbatim}

\newpage

\begin{center}
{\bf TEST RUN OUTPUT}
\end{center}

\begin{verbatim}

          16O   +           144Sm     Fusion reaction
-------------------------------------------------
Phonon Excitation in the targ.: beta_N= 0.205, beta_C= 0.205, r0= 1.06(fm)
                                      omega= 1.81(MeV), Lambda= 3, Nph= 1
-------------------------------------------------
Potential parameters: V0=  105.10(MeV), r0= 1.10(fm), a= 0.75(fm)
   Uncoupled barrier: Rb=10.82(fm), Vb=   61.25(MeV), Curv= 4.25(MeV)
-------------------------------------------------

       Ecm (MeV)    sigma (mb)          <l>
       -------------------------------------
       55.00000    0.97449E-02        5.87031
       56.00000        0.05489        5.94333
       57.00000        0.28583        6.05134
       58.00000        1.36500        6.19272
       59.00000        5.84375        6.40451
       60.00000       20.59856        6.86092
       61.00000       52.14435        7.81887
       62.00000       94.62477        9.18913
       63.00000      139.58988       10.65032
       64.00000      185.55960       11.98384
       65.00000      234.04527       13.13045
       66.00000      283.93527       14.18620
       67.00000      333.26115       15.21129
       68.00000      381.21017       16.20563
       69.00000      427.61803       17.16333
       70.00000      472.48081       18.08211
       71.00000      515.83672       18.96273
       72.00000      557.73621       19.80734

\end{verbatim}


\newpage
\begin{thebibliography}{99}

\bibitem{B88}M. Beckerman, Rep. Prog. Phys. 51 (1988) 1047; Phys. Rep.
129 (1985) 145.

\bibitem{DHRS98}M. Dasgupta, D.J. Hinde, N. Rowley, and A.M. Stefanini,
Annu. Rev. Nucl. Part. Sci. 48 (1998) 401.

\bibitem{BT98}A.B. Balantekin and N. Takigawa, Rev. Mod. Phys. 
70 (1998) 77. 

\bibitem{RSS91}N. Rowley, G.R. Satchler, and P.H. Stelson,
Phys. Lett. B254 (1991)25.

\bibitem{LDH95}J.R. Leigh, M. Dasgupta, D.J. Hinde, J.C. Mein,
C.R. Morton, R.C. Lemmon, J.P. Lestone, J.O. Newton, H. Timmers,
J.X. Wei, and N. Rowley, Phys. Rev. C52 (1995) 3151.

\bibitem{HTD97}K. Hagino, N. Takigawa, M. Dasgupta, D.J. Hinde,
and J.R. Leigh, Phys. Rev. C55 (1997) 276.

\bibitem{SACN95}A.M. Stefanini, D. Ackermann, L. Corradi, D.R. Napoli,
C. Petrache, P. Spolaore, P. Bednarczyk, H.Q. Zhang, S. Beghini,
G. Montagnoli, L. Mueller, F. Scarlassara, G.F. Segato, F. Sorame,
and N. Rowley, Phys. Rev. Lett. 74 (1995) 864.

\bibitem{HTBB95}K. Hagino, N. Takigawa, A.B. Balantekin, and J.R. Bennett,
Phys. Rev. C52 (1995) 286.

\bibitem{LR84}R. Lindsay and N. Rowley, J. Phys. G10 (1984) 805.

\bibitem{T87}O. Tanimura, Phys. Rev. C35 (1987) 1600; Z. Phys. A327
(1987) 413.

\bibitem{LP84}S. Landowne and S.C. Pieper, Phys. Rev. C29 (1984) 1352.

\bibitem{RMR78}P. Ring, H. Massmann, and J.O. Rasmussen,
Nucl. Phys. A296 (1978) 50.

\bibitem{MSR66}M.A. Melkanoff, T. Sawada, and J. Raynal, {\it Methods in
Computational Physics}, vol. 6 (1966)1.

\bibitem{KR93}M.W. Kermode and N. Rowley, Phys. Rev. C48 (1993) 2326.

\bibitem{BBH99}G.F. Bertsch, P.F. Bortignon, and K. Hagino, to be
published (nucl-th/9811030).

\bibitem{DV86}C.H. Dasso and G. Pollarolo, Phys. Lett. B155 (1985) 223;
C.H. Dasso and A. Vitturi, Phys. Lett. B179 (1986) 337.

\end{thebibliography}
\end{document}